\documentclass[11pt,halfparskip]{scrartcl}

\usepackage{mathpazo}

\usepackage{graphicx}
\usepackage[utf8]{inputenc} 
\usepackage{amsmath,amssymb,amsthm} 

\usepackage{graphicx}
\usepackage[square]{natbib}
\usepackage{url}
\usepackage{booktabs}
\usepackage{threeparttable}
\usepackage{subcaption}

\usepackage{tikz}
\usepackage[tikz]{mdframed}
\usepackage{colortbl,xcolor}
\usepackage[a4paper,margin=25mm]{geometry}
\usepackage[ruled,vlined,linesnumbered]{algorithm2e}

\usepackage{verbatim}
\usepackage{xspace}
\usepackage{rotating}

\usepackage{hyperref}

\newcommand{\ie}{ie}
\newcommand{\eg}{eg}
\newcommand{\reffig}[1]{Figure~\ref{#1}}

\definecolor{light-gray}{gray}{0.95}
\newmdenv[%
  outerlinewidth=2,%
  roundcorner=10pt,%
  linecolor=gray,%
  backgroundcolor=light-gray,%
]{myframe}

\usepackage{listings}
\usepackage{color}
\definecolor{codegreen}{rgb}{0,0.6,0}
\definecolor{codegray}{rgb}{0.5,0.5,0.5}
\definecolor{codepurple}{rgb}{0.58,0,0.82}
\definecolor{backcolour}{rgb}{0.95,0.95,0.92}
\lstdefinestyle{mystyle}{
    backgroundcolor=\color{backcolour},   
    commentstyle=\color{codegreen},
    keywordstyle=\color{magenta},
    numberstyle=\tiny\color{codegray},
    stringstyle=\color{codepurple},
    basicstyle=\footnotesize,
    breakatwhitespace=false,         
    breaklines=true,                 
    captionpos=b,                    
    keepspaces=true,                 
    numbers=left,                    
    numbersep=5pt,                  
    showspaces=false,                
    showstringspaces=false,
    showtabs=false,                  
    tabsize=2
}
\title{CityJSON: A compact and easy-to-use encoding of the CityGML data model}

\author{Hugo Ledoux \and Ken Arroyo Ohori \and Kavisha Kumar \and Balázs Dukai \and Anna Labetski \and Stelios Vitalis}

\date{} 

\begin{document}

\maketitle  

\begin{myframe}
This is the author's version of the work. 
It is posted here only for personal use, not for redistribution and not for commercial use.
The definitive version will be published in the journal \emph{Open Geospatial Data, Software and Standards} soon.
\\
\\
H. Ledoux, K. Arroyo Ohori, K. Kumar, B. Dukai, A. Labetski, and S. Vitalis (2018). CityJSON: A compact and easy-to-use encoding of the CityGML data model. \emph{Open Geospatial Data, Software and Standards}, In Press.
\\\textsc{doi}: \url{http://dx.doi.org/10.1186/s40965-019-0064-0}
\\
\\
Full details of the project: \url{https://cityjson.org}
\end{myframe}

\vspace{1cm}

\begin{abstract}
The international standard CityGML is both a data model and an exchange format to store digital 3D models of cities.
While the data model is used by several cities, companies, and governments, in this paper we argue that its XML-based exchange format has several drawbacks.
These drawbacks mean that it is difficult for developers to implement parsers for CityGML, and that practitioners have, as a consequence, to convert their data to other formats if they want to exchange them with others.
We present CityJSON, a new JSON-based exchange format for the CityGML data model (version 2.0.0).
CityJSON was designed with programmers in mind, so that software and APIs supporting it can be quickly built. 
It was also designed to be compact (a compression factor of around six with real-world datasets), and to be friendly for web and mobile development.
We argue that it is considerably easier to use than the CityGML format, both for reading and for creating datasets. 
We discuss in this paper the main features of CityJSON, briefly present the different software packages to parse/view/edit/create files (including one to automatically convert between the JSON and GML encodings), analyse how real-world datasets compare to those of CityGML, and we also introduce \emph{Extensions}, which allow us to extend the core data model in a documented manner.

\end{abstract}

%
\section{Introduction}%
\label{sec:intro}

CityGML is an open data model and exchange format to store digital 3D models of cities and landscapes, and it is standardised by the Open Geospatial Consortium~\citep{OGC-CityGML}. 
It defines ways to describe most of the common 3D objects found in cities (such as buildings, roads, rivers, bridges, vegetation and city furniture) and the (hierarchical) relationships between them.
It also defines different levels of detail (LoDs) for the 3D objects, allowing us to represent 3D city objects for different applications and purposes~\citep{15_ijgi_3dapps}.

As it can be observed from the CityGML specifications and the related scientific literature (for instance, among many, \citet{Groger12}, \citet{Kolbe08}, \citet{Kolbe08a}, \Citet{Brink13b}, \citet{Biljecki16c}), the vast majority of the efforts have been spent on developing the concepts and the data model.
In our opinion, very little attention has been paid to deriving a \emph{usable} exchange format.
Indeed, as we further explain in the paper, the only encoding that is standardised and supported by the OGC, an XML/GML-based one~\citep{OGC-GML}, is verbose, hierarchical, complex, and not adapted to the web.
We believe these drawbacks hinder the use of CityGML in practice, which can be observed by: (1) the low number of software packages supporting full read/write/edit capabilities for CityGML files; and (2) the relatively low number of datasets stored in CityGML files.

We present in this paper CityJSON (version 1.0.0), a JSON\footnote{The JavaScript Object Notation: \url{http://json.org}} encoding for the CityGML 2.0.0 data model.
JSON is, like GML, a text-based data exchange format that can be read both by humans and machines.
It was chosen as an alternative encoding to GML for several reasons.
First, and most importantly, JSON dominates the web: nowadays if two applications need to exchange data they will most likely use JSON (over XML).
According to \citet{Target17}, of the ten most popular APIs on the web, only one will expose its data in XML, the others all use JSON\@.
Second, JSON is predominantly favoured by developers (on \emph{Stack Overflow} it is by far the most discussed exchange format~\citep{Target17}) which means that more libraries and software will support it, and these will most likely be maintained.
Finally, JSON is based on two data structures that are available in virtually every programming language (more details below), and we can thus structure a file in the way that a developer would build and index in memory the objects (developers then do not need to use external libraries, all features and geometries are already indexed, and ready to use).

It should be observed that, at this moment, CityJSON is not an official OGC standard, and there are no concrete plans for it to become one.
It was developed to simplify the tasks of developers and thus to foster the use of the official data model in practice, but with a usable and simple-to-use encoding.

CityJSON follows the philosophy of another (non-standardised) encoding of CityGML: 3DCityDB~\citep{Yao18}.
That is, to be stored efficiently and allow practitioners to access features and their geometries easily, the deep hierarchies of the CityGML data model are removed and replaced by a `flat' representation.
Furthermore, there is one and only one way to represent the semantics and the geometries of a given feature, and some more additional restrictions are applied. 
The encoding of CityJSON allows us to bypass most of the drawbacks of the GML encoding: CityJSON files from real-world datasets are on average 6$\times$ more compact (we demonstrate this with real-world examples), and their structure can be parsed and manipulated easily by many programming languages, including JavaScript.
This can be seen in the easiness with which CityJSON software has been built so far (see section below).

CityJSON also supports extensions to the core data model of CityGML for specific applications and use-cases; in the CityGML world, these are called ADEs (application domain extensions) and several exist~\citep{Biljecki18}.
Our \emph{Extensions} are defined as simple JSON files, and support the addition of new feature types, as well as the addition of new attributes for features and for datasets.

%
\section{CityGML: a data model and an encoding}%
\label{sec:citygml}

\begin{figure}
  \centering
  \includegraphics[width=0.95\linewidth]{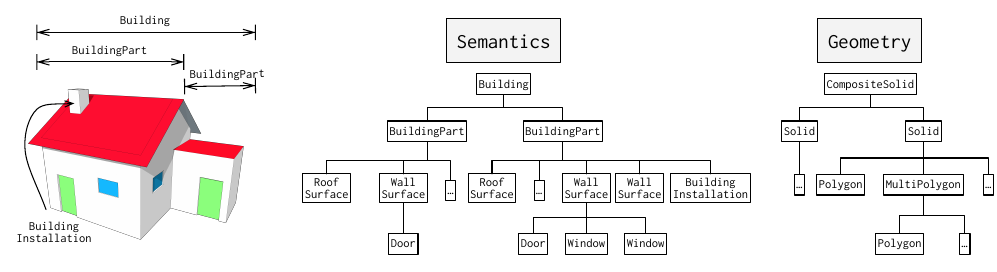}
  \caption{A building is semantically decomposed into different objects, and each objects is defined with geometry.}%
\label{fig:ssc}
\end{figure}

To represent a city, CityGML 2.0.0~\citep{OGC-CityGML,Groger12}\footnote{in the following it is assumed that CityGML refers to the latest version 2.0.0} recursively decomposes it into semantic objects.
It defines the classes most commonly found in an urban or a regional context, and the hierarchical relationships between them (\eg\ a building is composed of parts, which are formed of walls, which have windows). 
Figure~\ref{fig:ssc} shows how a given building, containing two parts, would be decomposed semantically and geometrically; notice that both decompositions should ideally be coherent~\citep{Stadler07}.
The geometry of the objects is realised with a subset of the geometry definitions in ISO 19107\footnote{Only linear and planar primitives are allowed.} \citep{ISO19107}, which also allows aggregations of geometries (Multi/Composite geometries): a single building can for instance be modelled with a \texttt{CompositeSolid}, such as that in Figure~\ref{fig:ssc}.
The CityGML semantic classes are structured into several modules, \eg\ Building, Land Use, Water Bodies, or Transportation.

One of the main characteristics of CityGML is that it supports five levels of detail (LoDs) for each of the classes.
This allows practitioners to use appropriate representations of a city depending on the application.

It is possible to extend the list of classes with new ones, and also to define new attributes.
The mechanism to accomplish this is called application domain extension (ADE), and involves creating new XML schemas that inherit from the CityGML XML schemas.

A CityGML file, encoded with XML, is structured in a hierarchy that ultimately reaches down to individual objects and their attributes. 
These objects have a geometry that is described using GML, and it is possible to attach textures and/or material to each of the surfaces.

\subsection{Main criticism of CityGML as an encoding}%
\label{sec:criticism}

CityGML files are known to be very difficult to parse and to extract information from.
We briefly describe the main issues, at three levels: XML, GML, and CityGML.

\paragraph{XML.} 
As mentioned in the Introduction, when JSON suffices, it is usually preferred by developers over XML\@.
This preference is mainly because JSON is far simpler than XML, which reflects the fact that JSON is a data format while XML is a markup language, and it is thus much easier to write software for JSON than XML\@.
Some common arguments in favour of JSON include its smaller file size, greater ease of reading (by humans), greater ease of parsing (by software), and the general difficulty of dealing with malformed XML (which is common)\@.
JSON is also based on simple data types and data structures that are available in almost all programming languages, and thus mapping the content of a file to a data structure that can be easily queried is trivial.
In fact, many programming languages treat JSON as a near-native type and can read and write to it (ie\ serialisation and deserialisation) without the need of external libraries.
By contrast, XML usually needs to be parsed in a process that requires the use of libraries and which creates a hierarchy of more complex objects, and this structure still needs to be traversed according to the logic of CityGML\@.

\paragraph{GML.} 
While GML allows us to represent geometries, the fact that there are many different ways to store the same geometry is a big handicap in practice.
A vivid example is shown by \citet{Rouault14}: a simple square can be stored in at least 25 different variations in GML\@.
The large number of possible variations means that a developer needs to figure out all possible variations for every geometry type and write code to handle them appropriately.
The variations also increase with the dimensionality, so the storage of a solid would have even more variations (because polygons are used as building blocks).
Apart from the large efforts required to find and handle all possible configurations, this is a situation that causes differences in how different software packages handle different datasets.

\paragraph{CityGML.} 
CityGML builds upon XML and GML, and so it inherits most of the advantages and disadvantages of these formats, but some specific features of CityGML are also problematic.
First, one thing to notice is that a city can be large and as a consequence CityGML files tend to be massive (1GB+ are common). 
As this means that they often do not fit in the memory of a computer, CityGML software sometimes needs to do more complex processing than it would be required otherwise (\eg\ dynamically reading from and writing to a database).
Second, as it can be seen in \reffig{fig:ssc}, the hierarchy for a single simple building can become rather deep, which translates into many classes which are nested hierarchically in XML\@.
This makes files very difficult to read by a person (which nullifies a big advantage of XML), and the many different classes might need specialised code to be written for each.
Third, CityGML makes extensive use of \emph{XLinks} (XML Linking Language).
While these are in theory powerful, in practice the links need to be resolved, which is problematic, especially for large files, or when references are external URIs (\ie\ pointing to objects not in the file).
Several XML libraries and software do not resolve XLinks.
Furthermore, many of the key features of CityGML are based on XLinks, \eg\ they are necessary for semantic surfaces.

The following are also problematic:
\begin{enumerate}
  \item semantic surfaces can be stored in many different ways (similar to GML versions of a polygon), a trait that is often seen in practice.
  \item \emph{Implicit Geometries} make extensive use of \emph{XLinks}, and one issue is that a given template feature can be located anywhere in a file. It is thus the burden of the developer to read the whole file, index all potential features, and then resolve them.
  \item because GML is used, the CRS of each object in a file can be defined. This means that all the objects in a file could in theory be of a different CRS\@. Even a building could have its windows defined in a different CRS\@. This means that a standard-compliant CityGML software needs to contain projection libraries.
  \item use of GML means that all 3D geometries are according to Simple Features, which means no topology is stored. 
\end{enumerate}

The consequences of the above is that software support for CityGML is lacking.
As a telltale example, there are still no full JavaScript parsers for CityGML (which are necessary in order to exchange and process files on the web), and thus the efficient exchange and processing of CityGML models on the web is very difficult, if not impossible.
We emphasise the word ``full'' here, given that there are existing JavaScript parsers but these are rather limited, as they are usually hard-coded for specific files or files written by a specific program.
When used with other files, they might therefore ignore some parts of a file or simply crash, because a particular CityGML representation has not been accounted for.

%
\section{CityJSON}%
\label{sec:cityjson}

The current version of CityJSON implements most of the CityGML data model, and all of the CityGML modules have been mapped.
The parts that were not implemented are based on the fact that they would have unnecessarily complicated the encoding, and that they are not used in practice (with the files that are publicly available at least).

We explain in the following the main engineering choices that were made, and we also describe where and how the data model differs from that of CityGML\@.
The full specifications are available online at \url{https://cityjson.org/specs/}.

The JSON data format defines simple data types for boolean values, numbers, and strings, as well as two data structures: 
\begin{enumerate}
  \item An ordered list of elements, which are separated by commas and enclosed with square brackets, \ie\ \texttt{[]}. We refer to it as an ``array''.
  \item An object consisting of key/value pairs (key is often named ``property''), which are in the form \texttt{key: value} and are enclosed with curly brackets, \ie\ \texttt{\{\}}. We refer to it as a ``dictionary''. It is often called a map, a hash table, an associative array, or in the context of JSON simply as an object.
\end{enumerate}

A JSON object can be any combination and nesting of the above elements.

A CityJSON file represents a given geographical area; the file contains one JSON object of type \texttt{"CityJSON"} and would typically contain the following JSON properties:
\begin{lstlisting}
  {
    "type": "CityJSON",
    "version": "1.0",
    "CityObjects": {},
    "vertices": [],
    "appearance": {}
  }
\end{lstlisting}

%
\subsection{City objects are ``flattened out''}

\begin{figure}
  \centering
  \includegraphics[width=0.6\linewidth]{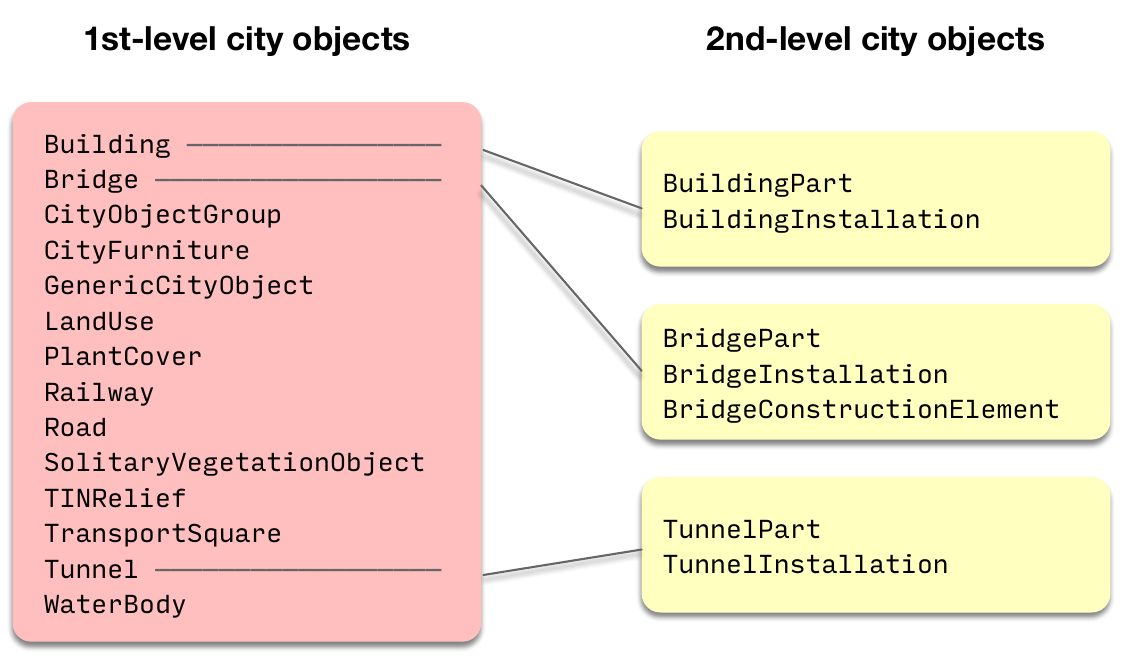}
  \caption{The implemented CityJSON classes (same name as CityGML classes) are divided into 1st and 2nd levels.}
\label{fig:co}
\end{figure}

The property \texttt{"CityObjects"} contains a dictionary where the properties are the identifiers of the city objects (\emph{IDs}).
The schema of CityGML has been flattened out and all hierarchies removed.
Figure~\ref{fig:co} shows the city objects that are supported in CityJSON, both 1st- and 2nd-level city objects are stored in the dictionary \texttt{"CityObjects"}.

As an example, for a building containing 2 parts, the 3 objects will be represented at the same level and linked by their \emph{IDs}.
\begin{lstlisting}
  "CityObjects": {
    "id-1": {
      "type": "Building",
      "attributes": {...},
      "children": ["id-2", "id-3"],
      "geometry": [{...}]
    },
    "id-2": {
      "type": "BuildingPart",
      "parents": ["id-1"],
      "geometry": [{...}]
      ...
    },
    "id-3": {
      "type": "BuildingPart",
      "parents": ["id-1"],
      "geometry": [{...}]
      ...
    }
  }
\end{lstlisting}

Each city object can have a \texttt{"parents"} and/or a \texttt{"children"} property, and this is how in the snippet the building \texttt{"id-1"} is linked to its 2 parts.
The fact that a dictionary is used means that developers have direct access to the city objects through their IDs (and also in constant time if a hashmap is used to implement the dictionary).

A city object can be of any of the types defined in Figure~\ref{fig:co}, and each of them must have the same structure, and at a minimum contain a \texttt{"geometry"} property. 
If attributes are to be stored, they have to be in the \texttt{"attributes"} property.
This simplifies the work of the developer because there is a single point of entry for all geometries and attributes, unlike with CityGML.
\begin{lstlisting}
  {
    "type": "PlantCover",
    "attributes": {
      "averageHeight": 11.05,
      "colour": "green"
    },
    "geometry": [{...}]
  }
\end{lstlisting}

%
\subsection{Geometry}

CityJSON defines the same 3D geometric primitives used in CityGML, with the same restrictions for linearity/planarity.
However, since they are rarely used in a 3D context, \emph{Point} and \emph{LineString} only have their Multi* counterparts; a single \emph{Point} is a \emph{MultiPoint} with only one object.
When a geometry is defined, it must contain a value for the LoD. 
In order to avoid ambiguities, we encourage the use of the refined LoDs, as defined in \citet{Biljecki16}, over the five standard CityGML ones.
City Object can have several LoDs, and thus CityJSON, as is the case for CityGML, allows us to store concurrently several LoDs for the same object.
\begin{lstlisting}
  {
    "type": "MultiSurface",
    "lod": 2.1,
    "boundaries": [
      [[0, 3, 2, 1]], [[4, 5, 6, 7]], [[0, 1, 5, 4]]
    ]
  }
\end{lstlisting}

It should be noticed that CityJSON uses a different approach from (City)GML to store the $(x,y,z)$ coordinates of geometric primitives.
A geometric primitive does not list all the coordinates of its vertices, rather the coordinates of the vertices are stored in a separate array (the \texttt{"vertices"} property of the CityJSON object), and geometric primitives refer to the position of a vertex in that array.
\begin{lstlisting}
  "vertices": [
    [8623.234, 487111.009, 13.92],
    [8829.456, 488115.134, 10.07],
    [8554.508, 487229.995, 19.61],
    ...
    [8523.134, 487625.134, 2.03]
  ]
\end{lstlisting}
The indexing mechanism of the format \emph{Wavefront OBJ}\footnote{\url{https://en.wikipedia.org/wiki/Wavefront_.obj_file}} is reused, because it has been used for many years, with success, in the computer graphics community.
There are several advantages to this approach.
First, the files can be compressed: 3D vertices are often shared by several surfaces, and repeating them can be costly (especially if they are very precise, often sub-millimetre is used).
Second, this increases the topological relationships that are explicitly stored in the file, and several operations can be sped up and made more robust (\eg\ are two buildings adjacent?).
Third, it is very easy to convert to a representation listing all coordinates; the inverse is not true.

The geometry is based on an enumeration of the vertices forming each ring of a surface, as follows.
A \texttt{"MultiSurface"} has an array containing surfaces, where each surface is modelled by an array of arrays, the first array being the exterior boundary of the surface, and the others the interior boundaries.
A \texttt{"Solid"} has an array of shells, the first array being the exterior shell of the solid, and the others being the interior shells; each shell has an array of surfaces, modelled in the exact same way as a \texttt{"MultiSurface"}.
Notice that unlike with (City)GML, there is only one variation per geometry type, which (greatly) simplifies the life of developers.
\begin{lstlisting}
  {
    "type": "Solid",
    "lod": 2.2,
    "boundaries": [
      [ [[0, 3, 2, 1, 22]], [[4, 12, 123, 5, 6, 7]], [[0, 1, 5, 4]], [[1, 2, 6, 5]] ], //-- exterior shell
      [ [[240, 243, 124]], [[244, 246, 724]], [[34, 414, 45]], [[111, 246, 5]] ] //-- interior shell
    ]
  }
\end{lstlisting}

%
\subsection{Semantic surfaces}

In one given city object (say a \texttt{"Building"}), several surfaces can have the same semantics (think for instance of a complex building that has been triangulated, there can be many triangles for one given surface).
Because of this, a semantic surface, which is a pivotal concept in CityGML, becomes a JSON object that is stored separately from the geometry of a city object.
By doing so, a semantic surface object has to be declared only once, and each of the surfaces used to represent it can point to it. 
This is achieved by first declaring all the semantic surfaces in a \texttt{"surfaces"} array, and then declaring an added \texttt{"values"} array that links each surface to its corresponding semantic surface using their respective positions in the arrays.
\begin{lstlisting}
  {
    "type": "Solid",
    "lod": 2,
    "boundaries": [
      [ [[0,3,2,1,22]], [[4,5,6,7]], [[0,1,5,4]], [[1,2,6,5]] ]
    ],
    "semantics": {
      "surfaces" : [
        { "type": "RoofSurface" },
        {
          "type": "WallSurface",
          "paint": "blue"
        },
        { "type": "GroundSurface" }
      ],
      "values": [ [0, 1, 1, 2] ]
    },
  }
\end{lstlisting}

%
\subsection{Geometry templates}
\label{sec:templates}

CityGML's \emph{Implicit Geometries}, better known in computer graphics as \emph{templates}, are one method to compress files since identical geometries (\eg\ benches, lamp posts, and trees), need only be defined once (and translations/rotations/scaling are applied).
In CityJSON, they are implemented slightly differently than in CityGML: they are stored at one specific location in the file, and each template can be reused. 
In CityGML, one reuses the geometry used for another city object, and thus there is no structured way to store them, and furthermore, one has to search for them in the file (with XLinks) because they can be located anywhere (the link could even point to an external reference that needs to be resolved).
\begin{lstlisting}
  "geometry-templates": {
    "templates": [
      {
        "type": "MultiSurface",
        "lod": 2,
        "boundaries": [ 
           [[0, 3, 2, 1]], [[4, 5, 6, 7]], [[0, 1, 5, 4]]
        ]
      }
    ],
    "vertices-templates": [...]
  }
\end{lstlisting}
A given city object can have a geometry of type \emph{"GeometryInstance"} (instead of those defined above), which defines the ($x,y,z$) location, a link to the geometry template, and the transformation matrix.
\begin{lstlisting}
  {
    "type": "SolitaryVegetationObject", 
    "geometry": [
      {
        "type": "GeometryInstance",
        "template": 0,
        "boundaries": [372]
        "transformationMatrix": [
          2.0, 0.0, 0.0, 0.0,
          0.0, 2.0, 0.0, 0.0,
          0.0, 0.0, 2.0, 0.0,
          0.0, 0.0, 0.0, 1.0
        ]
      }
    ]
  }
\end{lstlisting}

%
\subsection{Appearance}

Both textures and materials are supported, and the same mechanisms as CityGML are used for these. 
The material is represented with the X3D specifications\footnote{\url{https://en.wikipedia.org/wiki/X3D}}, as is the case for CityGML\@. 
For the texture, the COLLADA specifications\footnote{\url{https://www.khronos.org/collada/}} are reused, as is the case for CityGML\@.

Just as for the geometry templates, all material and textures must be located at the same entry point in a CityJSON file; this is in contrast to CityGML where they can be located anywhere.

%
\subsection{Schema validation}

CityJSON uses schemas defined in JSON Schema\footnote{\url{https://json-schema.org/}} to document its data model and to validate whether a CityJSON file respects the allowed structure and syntax.
All the city objects, their attributes, the allowed geometries, and other constraints are defined in schemas that are openly available at \url{https://cityjson.org/schemas/}.

It should be noticed that JSON Schemas are less flexible than XML Schemas, inheritance and namespaces are for instance not supported.
They nevertheless allow us to document most of what is possible with XML, and we have added extra validation functions to the software \emph{cjio} for the properties and constraints that cannot be expressed with JSON Schemas, see the section about software below for details.
The extra constraints can be seen as validating the internal consistency of a given CityJSON file, and examples of these are:
\begin{itemize}
  \item are the links between 1st- and 2nd-level city objects consistent?
  \item are the arrays for the boundaries and the semantics coherent? (\ie\ same structure)
  \item are there duplicate \emph{IDs} for city objects? 
  \item are there duplicate or orphan vertices?
  \item are there vertex indices that do not exist? 
\end{itemize}

%
\subsection{CityGML support}

CityJSON implements most of the data model, and all the CityGML modules have been mapped to CityJSON objects. 
However, for the sake of simplicity and efficiency, some modules and features have been omitted and/or simplified. 
If a module is supported, it does not mean that there is a 1-to-1 mapping between the classes and features in CityGML and CityJSON, but rather that it is possible to represent the same information, but in a different manner. 
CityJSON is thus conformant to a subset of CityGML, although technically only CityGML files (encoded with the XML format) can be conformant to the specifications of CityGML~\citep[Clause~2 about Conformance]{OGC-CityGML}.

The main features that are \underline{not} supported are:
\begin{itemize}
  \item The LoD4 of CityGML, which was mostly designed to represent the interior of buildings (including details and furniture), is not implemented. The main reason is that this concept will be revamped completely in the next CityGML version~\citep{Lowner16}, and currently there are virtually no datasets having LoD4 buildings.
  \item No support for arbitrary coordinate reference systems (CRSs). Only an EPSG code\footnote{\url{https://epsg.io}} can be used. 
  \item All geometries in a given CityJSON object must use the same CRS\@.
  \item In CityGML most objects can have an ID (usually \texttt{gml:id}). That is, not only can one building have an ID, but also each 3D primitive forming its geometry can have an ID\@. In CityJSON, only city objects and semantic surfaces can have IDs.
\end{itemize}

%
\subsection{Compression of CityJSON files}
\label{sec:compression}

To reduce the size of a file, it is possible to represent the coordinates of the vertices with integer values, and store the scale factor and the translation needed to obtain the original coordinates (stored with floats/doubles).
If compressed, a CityJSON file contains a \texttt{"transform"} property:
\begin{lstlisting}
  "transform": {
      "scale": [0.01, 0.01, 0.01],
      "translate": [4424648.79, 5482614.69, 310.19]
  }
\end{lstlisting}
and the real-world coordinates of a given vertex $v$ are obtained easily, for example for the $x$ component:
\[
  x = (v_x * transform.scale_x) + (transform.translate_x)
\]

Several file formats use this, for instance LAS~\citep{LAS-specs} and TopoJSON~\citep{Bostock18}.
For CityJSON, it typically compresses the files by around 5--10\%; we give below examples with  real-world datasets.
It should be noticed that it also makes files more ``robust'', in the sense that the coordinates are not prone to rounding because of floating-point representation in a computer~\citep{Goldberg91}.
This is the favoured way to store CityJSON files.

%
\subsection{Handling and streaming (large) CityJSON files}

One drawback of representing geometries by having references to a list of vertices is that large files are difficult to handle (one needs to read all of the file in memory to reconstruct the geometries) and that streaming of large files is thus complicated.

There exists a misconception that CityGML, since it uses the Simple Features paradigm~\citep{OGC-SF}, can be easily and directly streamed. 
We claim that while it is easier, this is not completely true.
CityGML files also often contain references between objects in a given file (XLinks), and before this file can be streamed, these references need to be resolved and the objects copied to the location pointing to it.
This also increases the size of the file.

\citet{Isenburg05} proposes to reorganise the order of the information in the file so that the vertices are not all at the end, they rather are located close to the geometries that need them.
Special tags in the file informs us about the fact that a vertex will not be used anymore, thus allowing us to free the memory.

This cannot be used with the current structure of CityJSON, but we propose instead to partition a CityJSON file into several files.
The rule can be based on a spatial partition, on the type of city objects, or simply randomly.
It suffices to update the list of vertices and the indices, which is a simple operation.
The open-source software cjio has an implementation of this.

Partitioning a given CityJSON file into several usually will not increase the storage.
There will be several properties (\eg\ the CRS, metadata, etc.) that will be repeated for each of the files, but the indices in each file will be smaller (always starting at 0), and thus in practice we have noticed that the size will actually \emph{decrease}.

%
\subsection{Support for metadata}

CityGML has very limited support for metadata~\citep{18_ogdss_metadata}. 
Only a few elements are supported, such as the bounding box and the CRS, and most elements are on the city model level and not on the module or city feature level. 
While there exists a metadata ADE for CityGML\footnote{\url{https://github.com/tudelft3d/3D_Metadata_ADE}}, in CityJSON metadata is incorporated into the core schema. 
CityJSON metadata is developed with ISO 19115 (the metadata standard specifically for geographic information developed by the International Organization for Standardization) as the base and further includes elements important for 3D city models, such as the levels of detail present, extensions (and their metadata), presence of textures and/or materials, etc. 
It also supports metadata at the city model level, the module level and the city feature level.

This is the only addition that CityJSON makes to the CityGML data model.

%
\section{Implementation and experiments}%
\label{sec:implementation}

\subsection{Software to read/write CityJSON}
\label{sec:software}

There are already several software programs to create, parse, visualise, and edit CityJSON files.
These were written in different languages (mostly Java, C++, Python, and Ruby) and have been coded during the development of the CityJSON specifications; our workflow involved testing new features to ensure that in practice they are implementable.

The structure of a CityJSON file has been developed so that the developer who wants to parse the file does not have to use an auxiliary data structure to index and extract information from the file.
One example is that all city objects are indexed in a dictionary (by their \emph{identifier}), which allows the developer to have direct access to them; this is particularly useful because the city objects have been flattened out, and a \texttt{"Building"} refers to its \texttt{"BuildingPart"}s by their identifiers.
Many other features of CityJSON are based on the simple indexing of objects in an array (templates, textures, materials, etc.), and thus they can be accessed directly by their index in the array.

We provide in this section an overview of a few software implementations, but this list is not exhaustive.

\paragraph*{citygml4j:}
an open-source Java class library and API for facilitating the reading/writing/editing of CityGML files.
Starting from version 2.6.0, it supports parsing and writing CityJSON, and all of the features of CityJSON are supported.
It can automatically convert CityGML to CityJSON (and vice-versa); the datasets used for the experiments in this paper have all been automatically converted with citygml4j. 
[\url{https://github.com/citygml4j/citygml4j}]

\paragraph*{cjio:}
a Python command-line interface (CLI) program to process and manipulate CityJSON files. 
The different operators can be chained to perform several processing operations in one step (thus avoiding saving several temporary files).
Examples of operators are: creating a subset given certain rules, validating with the CityJSON schemas, merging several files in one, reprojecting to a different CRS, and modifying the paths for the textures. 
[\url{https://github.com/tudelft3d/cjio}]

\paragraph*{azul:}
a modern 3D viewer for macOS, written in Swift and C++.
It supports the primitives and semantical surfaces in CityJSON; textures, material and geometry templates are currently not supported.
The datasets in Figure~\ref{fig:datasets} are visualised in azul.
[\url{https://github.com/tudelft3d/azul}]

\paragraph*{3dfier:}
a software to automatically construct 3D city models from 2D GIS datasets and elevation datasets (LiDAR).
The polygons are lifted to their elevation, and their semantics is taken into account.
One of the output formats of 3dfier is CityJSON\@.
[\url{https://github.com/tudelft3d/3dfier}]

\paragraph*{QGIS plugin:}
a simple QGIS plugin to load CityJSON files has been developed in Python.
The city objects are loaded as features in layers and can be divided and styled in different layers according to their object type; their geometry can be visualised both in the 2D and 3D view, while their semantic information can be displayed in the attribute table.
[\url{https://github.com/tudelft3d/cityjson-qgis-plugin}]

\paragraph*{val3dity:}
a validator for the 3D geometries defined in ISO 19107~\citep{ISO19107}. Written in C++. CityJSON fully supported.
Full details of the implementation in \citet{Ledoux13b} and \citet{18_ogdss_val3dity}.
[\url{https://github.com/tudelft3d/val3dity}]

\paragraph*{CityJSON web-viewer:}
a simple web-based viewer written in JavaScript.
Anyone can simply open a local file and visualise it, all the operations are done locally in the browser.
It does not support attributes querying or other queries at this moment, but demonstrates that simple tools can be built quickly if the encoding is simple.
[\url{https://tudelft3d.github.io/CityJSON-viewer}]

\subsection{Experiments with real-world datasets}%
\label{sec:experiments}

\begin{figure}
  \centering
  \begin{subfigure}[b]{0.3\linewidth}
    \centering
    \includegraphics[width=\textwidth]{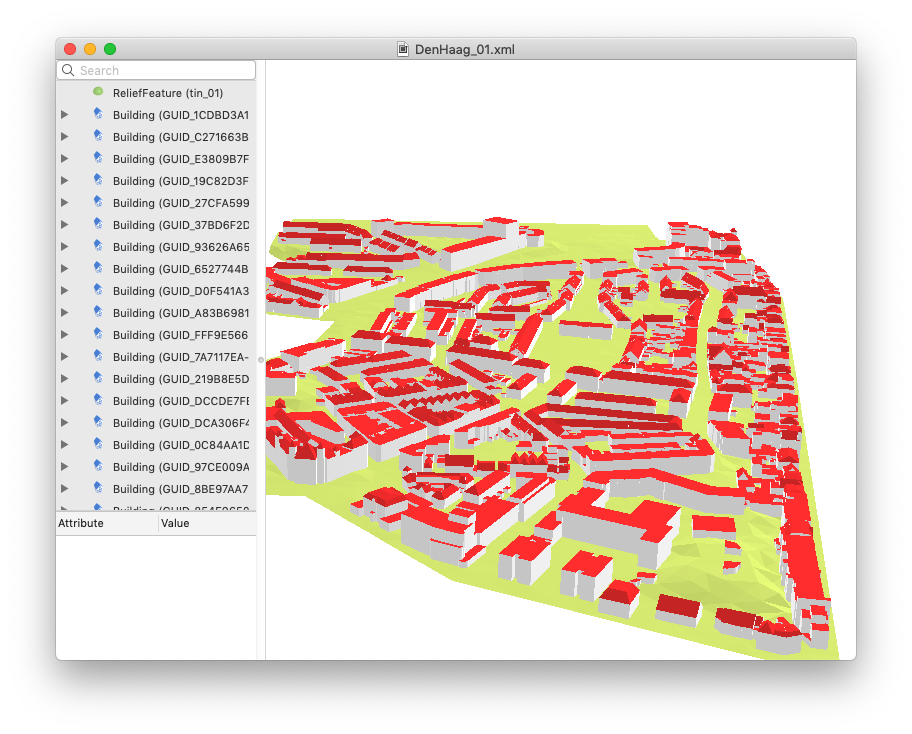}
    \caption{Den Haag}
  \end{subfigure}%
  \begin{subfigure}[b]{0.3\linewidth}
    \centering
    \includegraphics[width=\textwidth]{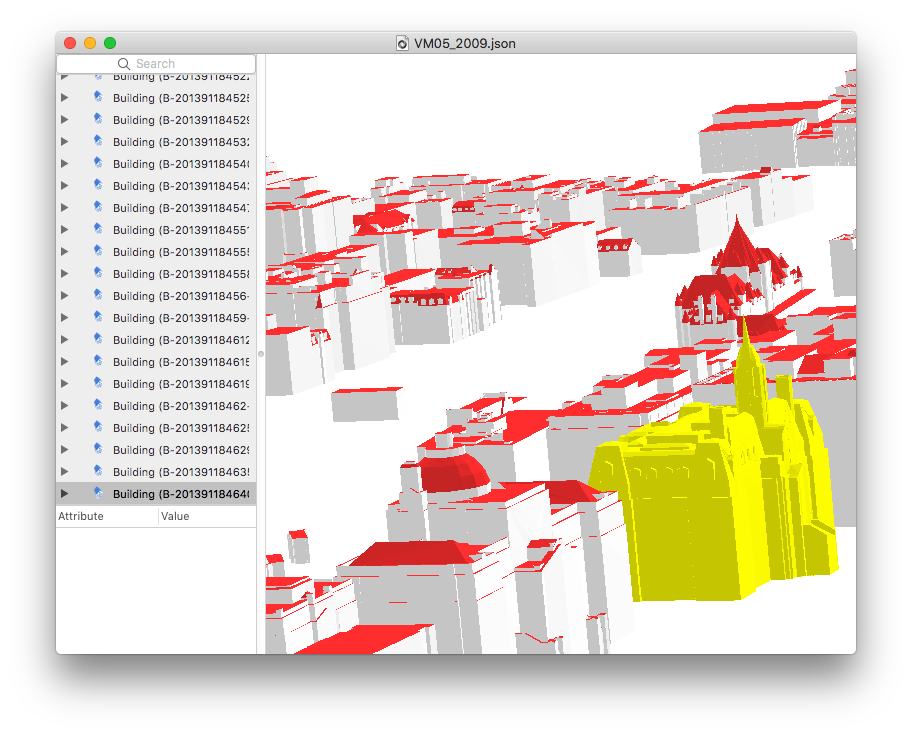}
    \caption{Montréal}
  \end{subfigure}%
  \begin{subfigure}[b]{0.3\linewidth}
    \centering
    \includegraphics[width=\textwidth]{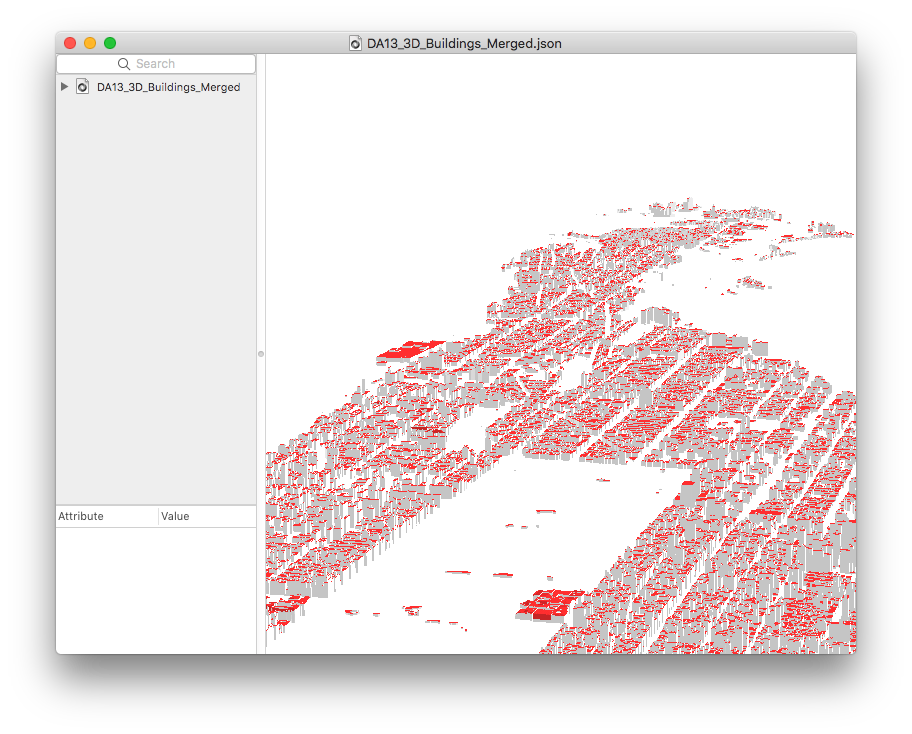}
    \caption{New York}
  \end{subfigure}
  \begin{subfigure}[b]{0.3\linewidth}
    \centering
    \includegraphics[width=\textwidth]{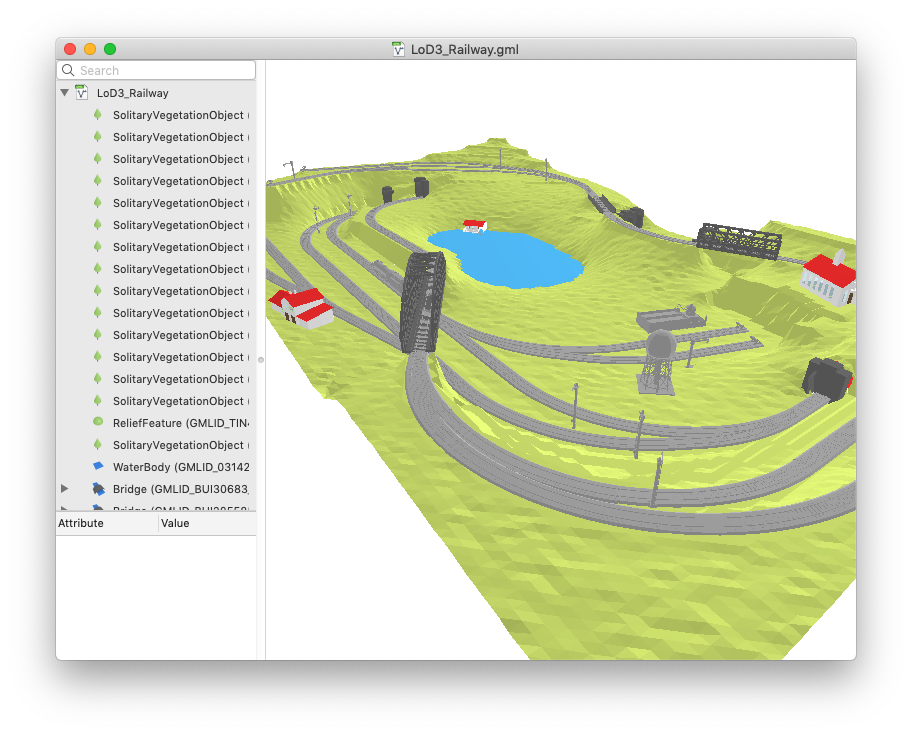}
    \caption{Railway}
  \end{subfigure}%
  \begin{subfigure}[b]{0.3\linewidth}
    \centering
    \includegraphics[width=\textwidth]{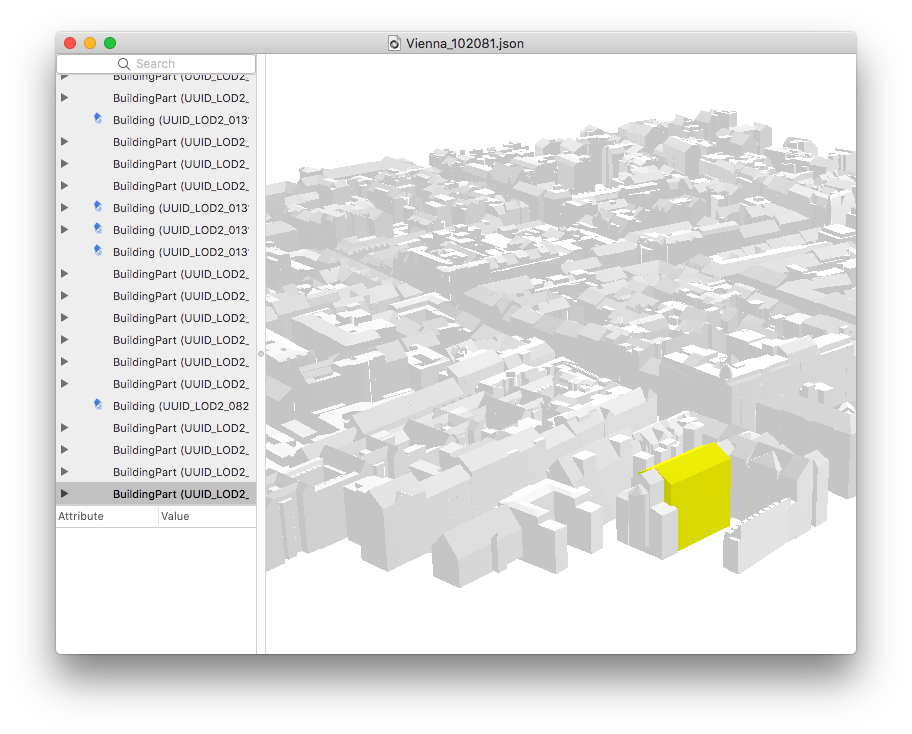}
    \caption{Vienna}
  \end{subfigure}
  \begin{subfigure}[b]{0.3\linewidth}
    \centering
    \includegraphics[width=\textwidth]{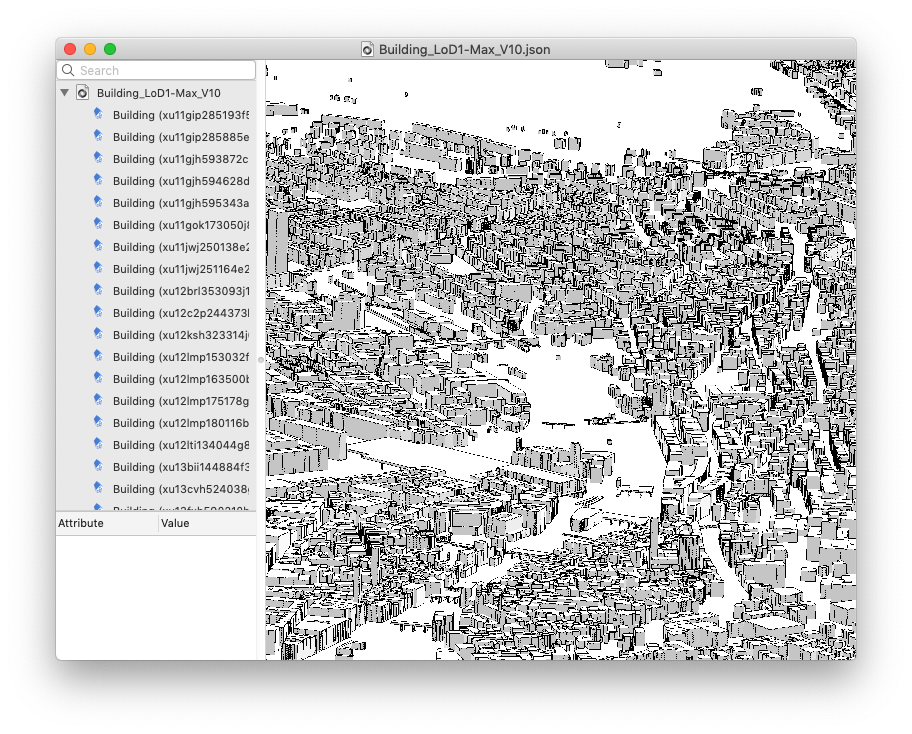}
    \caption{Zürich}
  \end{subfigure}%

\caption{CityGML datasets used for the experiments, details in Table~\ref{tab:datasets}.}
\label{fig:datasets}
\end{figure}

To demonstrate and test the software packages mentioned above, we have taken a few subsets of openly available datasets stored in CityGML, and converted them automatically (with citygml4j) to CityJSON.
The datasets used are shown in Table~\ref{tab:datasets} and Figure~\ref{fig:datasets}.
\begin{table}
  \centering
  \begin{threeparttable}
  \caption{Datasets converted (see Figure~\ref{fig:datasets}).}
  \label{tab:datasets}

  \begin{tabular}{@{}lrrrrrrrrrr@{}} \toprule
    &&  \multicolumn{4}{c}{CityGML} && \multicolumn{2}{c}{CityJSON} &  \\ 
    \cmidrule{3-6} \cmidrule{8-9} 
    && size\footnotesize ${}^{\text{(a)}}$ & no space\footnotesize ${}^{\text{(b)}}$ & LoD & texture && size-float\footnotesize ${}^{\text{(c)}}$ & size-int\footnotesize ${}^{\text{(d)}}$ && compr.\footnotesize ${}^{\text{(e)}}$ \\
    \midrule

    \textbf{Den Haag} \footnotesize ${}^{\text{(1)}}$
    && 23MB & 18MB & 2 & material && 3.1MB & 2.9MB && 6.2 \\

    \textbf{Montréal} \footnotesize ${}^{\text{(2)}}$ 
    && 56MB & 42MB & 2 & yes && 5.7MB & 5.4MB && 7.8 \\

    \textbf{New York}\footnotesize ${}^{\text{(3)}}$
    && 590MB & 574MB & 2 & no && 110MB & 105MB && 5.5 \\

    \textbf{Railway}\footnotesize ${}^{\text{(4)}}$ 
    && 45MB & 34MB & 3 & yes && 4.5MB & 4.3MB && 8.1 \\

    \textbf{Vienna} \footnotesize ${}^{\text{(5)}}$
    && 37MB & 36MB & 2 & no && 5.6MB & 5.3MB && 6.8 \\

    \textbf{Zürich} \footnotesize ${}^{\text{(6)}}$ 
    && 435MB & 423MB & 1 & no && 127MB & 100MB && 4.4 \\

    \bottomrule
  \end{tabular}
  \begin{tablenotes}[flushleft]
    \footnotesize
    \item ${}^{\text{(a)}}$ size does not take into account the size of the textures files (PNG, JPG, etc) since CityJSON refers to the same ones
    \item ${}^{\text{(b)}}$ the carriage returns, tabs, and spaces are removed, for a fair estimation of the compression factor
    \item ${}^{\text{(c)}}$ coordinates represented as double/float
    \item ${}^{\text{(d)}}$ coordinates represented as integer (compressed files)
    \item ${}^{\text{(e)}}$ compression factor = CityGML(no spaces) / CityJSON(size-int)
    \item ${}^{\text{(1)}}$ tile \emph{01}, \url{https://data.overheid.nl/data/dataset/ngr-3d-model-den-haag}
    \item ${}^{\text{(2)}}$ tile \emph{VM05}, \url{https://tinyurl.com/y8eglpmn}
    \item ${}^{\text{(3)}}$ LoD2 tile \emph{DA13}, \url{https://www1.nyc.gov/site/doitt/initiatives/3d-building.page}
    \item ${}^{\text{(4)}}$ CityGML v2 demo \emph{Railway}, \url{https://www.citygml.org/samplefiles/}
    \item ${}^{\text{(5)}}$ the demo file, \url{https://tinyurl.com/yaopvy6w}
    \item ${}^{\text{(6)}}$ version with \emph{Max} height, \url{https://data.stadt-zuerich.ch/dataset/geo_3d_blockmodell_lod1}
  \end{tablenotes}
  \end{threeparttable}
\end{table}

These datasets were reconstructed with different methodologies and utilising different software, and they cover a wide-range of possibilities: textures, no textures, material, geometry templates, different LoDs, etc.

The first thing to notice is that CityGML files, as downloaded, often contain several carriage returns, extra spaces, and tabs, and these can significantly increase the file size (and have no use for computers).
We have therefore removed all of these to provide a fair comparison of the file size with CityJSON (which do not contain any either).
While this might seems unimportant, one can observe that for the CityGML datasets we used, the compression obtained is already large, \eg\ for the \emph{Montréal} dataset we obtain 25\%.

Notice also that the compression obtained by encoding the vertices in integers (see the section about compression of CityJSON files, above) can also be significant.

If we compare CityGML files without any spaces or carriage returns to the CityJSON files (integer coordinates), the average compression factor is about six (it varies between 4.4 and 8.1).
This varies because of several reasons: (1) if several geometries are shared/adjacent, in CityJSON the vertices are merged; (2) generic attributes are very verbose in CityGML, and in CityJSON they do not occupy extra space, they are considered simply as an attribute; (3) simple files like Zürich contain only simple LoD1 blocks, and there is no semantics or any other features used (\eg\ geometry templates); (4) some of the compression is obtained because in the original CityGML file each polygon has a \texttt{gml:id}, and this is lost during the translation (we believe this ID has little meaning in practice, and is stored simply because the export function created it).

Observe that the presence of textures and materials does not seem to affect the compression factor, this is explained by the fact that the sizes of the textures is not taken into account (both CityGML and CityJSON simply refer to the files on disk), and more or less the same mechanism is used. 

We have tried with several other different datasets, and we have obtained similar results.

%
\section{Extensions to the core data model}%
\label{sec:extensions}

The CityGML data model allows us to represent the most common city objects, but sometimes practitioners may want to model additional objects and/or add certain attributes to the data model.
For this, CityGML has the concept of ADEs (application domain extensions).
An ADE is defined in an extra XML Schema (XSD file) with its own namespace. 
Commonly, inheritance is used to refine the classes of the CityGML data model to define entirely new classes, and to modify any class by adding for instance new geometries and complex attribute~\Citep{Brink13}.
An ADE allows us to document in a structured way, and also to validate, an instance of a CityGML document that would contain both classes from the core model and from the ADEs.
There exists several ADEs, see \citet{Biljecki18} for an overview.

In a similar manner, CityJSON defines \emph{Extensions}.
An Extension is a JSON file that documents how the core data model of CityJSON may be extended, and is utilised in the validation of CityJSON files.
Unlike ADEs where the user is allowed to extend the data model in any way she wants, CityJSON restricts the possible cases to these three:
\begin{enumerate}
  \item Adding new complex attributes to existing city objects
  \item Creating a new city object, or ``extending'' one, and defining complex geometries
  \item Adding new properties at the root of a document
\end{enumerate}
While Extensions are less flexible than CityGML ADEs (inheritance and namespaces are for instance not supported, and less customisation is possible), it should be noted that the flexibility of ADEs comes at a price: the software processing an extended CityGML file will not necessarily know what structure to expect and how to handle it, which means that software support for them will likely be inconsistent.
There is ongoing work to use the ADE schemas to automatically do this~\citep{Yao17,Yao18}, but this currently is not supported by most software. 
Viewers might not be affected by ADEs because the geometries are usually not changed by an ADE\@. However, software parsing the XML to extract attributes and features might not work directly (and thus specific code would need to be written). 

Because Extensions cannot have namespaces (a limitation of JSON Schemas), to avoid conflicts between different Extensions we recommend prepending new City Objects and attributes with the name of the Extension; the lack of namespaces does not cause any other issues in practice.

A CityJSON Extension is a JSON file such as this one:
\begin{lstlisting}
  {
    "type": "CityJSON_Extension",
    "name": "Noise",
    "uri": "https://someurl.org/noise.json",
    "version": "0.1",
    "description": "Extension to model the noise"
    "extraRootProperties": {},     
    "extraAttributes": {},
    "extraCityObjects": {}
  }
\end{lstlisting}
It must define the name of the Extension, its URI, and its version.
The three cases to extend the core model, as described above, are three properties of the file.
Each of these properties contain snippets of JSON schemas, and these can reuse and refer to the definitions and geometric primitives defined in the schemas of CityJSON\@.

Since the file is not technically a JSON Schema file, there needs to be a software that preprocesses the file (or potentially other Extensions, since a given CityJSON file could contain several Extensions) and `links' it to the CityJSON definitions.
One of such software is \emph{cjio}.

CityJSON Extensions are designed such that they can be read and processed by standard CityJSON software without extra work on the developer's part.
Often no changes in the parsing code is required.
This is achieved by enforcing a set of simple rules when adding new city objects. 
If these are followed, then a CityJSON file containing Extensions will be seen as a `standard' CityJSON file.
Examples of these rules are:
\begin{enumerate}
  \item The name of a new city object must begin with a \texttt{+}, \eg\ \texttt{"+NoiseBarrier"}
  \item A new city object must conform to the rules of CityJSON, \ie\ it must contain a property \texttt{"type"} and one \texttt{"geometry"}. If the object contains appearances, the same mechanism should be used so that the new city objects can be processed without modification. 
  \item All the geometries must be in the property \texttt{"geometry"}, and cannot be located somewhere else deep in a hierarchy of a new property. This ensures that all the code written to process, manipulate, and view CityJSON files will be working without modifications. 
\end{enumerate}

As a concrete example, here is a snippet of the Extension in which we want to add two new attributes to the city object \texttt{"Building"}.
Both attributes start with a \texttt{"+"}, which is the CityJSON convention to add new objects and attributes. 
The first attribute is simply of type string, and the second one is a complex type to store a measurement.
\begin{lstlisting}
  "extraAttributes": {
    "Building": {
      "+noise-buildingReflection": { "type": "string" },
      "+noise-buildingReflectionCorrection": { 
        "type": "object",
        "properties": {
          "value": { "type": "number" },
          "uom": { "type": "string" }
        }
      }
    }
  }
\end{lstlisting}
A CityJSON file in which this Extension is used would look like this:
\begin{lstlisting}
  {
    "type": "CityJSON",
    "version": "1.0",
    "extensions": {
      "Noise": {
        "url" : "https://someurl.org/noise.json",
        "version": "0.1"
      }
    },
    "CityObjects": {
      "id-1234": {
        "type": "Building",
        "attributes": {
          "roofType": "gable",
          "+noise-buildingReflectionCorrection": {
            "value": 4.123,
            "uom": "dB"
          },
          "+noise-buildingRelection": { "facade" }
        },
        "geometry": [...]
      }
    }
  }
\end{lstlisting}

%
\section{Conclusions}%
\label{sec:discussion}

In programming, choosing the \emph{least} powerful language suitable for a given purpose is known as a principle of good design~\citep{BernersLee01}.
While doing so might seem limiting at first, it ultimately results in software and standards that are easier to design, write, test, and use.

In the context of 3D city models, we recognise that having an open standardised data model like CityGML is essential, but we also observe that its GML encoding can be overly complex in a way that is often unfriendly to developers.
The difficulties of parsing CityGML files, interpreting all the different ways in which geometries can be stored, resolving XLinks, dealing with different CRSs, and implementing support for ADEs, add up and create a high barrier for developers to support CityGML\@.
This discourages the adoption of the standard by developers and is especially hostile to small independent ones, such as those at the heart of the open-source GIS community.
This results in poor software support, and a lack of tools to do even basic processing (as can be currently observed in practice).

CityJSON greatly reduces the complexity of developing applications for the CityGML data model through the use of a simpler JSON encoding.
JSON is designed as a simple data interchange format and is natively supported by many programming languages, including JavaScript, Python, and Ruby.
Easy to use libraries add native-like support for it in many other popular languages, including C++.
Parsing a (City)JSON file is thus often a one-line operation that results in a tree of native data types, which can then be easily queried using standard functions.
In contrast, the developers who work with CityGML are often forced to write their own CityGML parser based on generic XML parsing libraries, which is a much more complex and error-prone process.
This is true even for simple operations, such as to assess if a file is fit-for-use within a specific application.

In recognition of the fact that a 3D city model format is of little us unless implemented, the development of the specifications of CityJSON has been done in a developer-centred process.
Each iteration of the specifications has been tested by implementing support for it in a few software packages with different programming languages.
By doing so, we were able to use the insight gained through this process to propose improvements for the next iteration, as well as to avoid the escalation of complexity that often occurs in geoinformation standards.
Moreover, since we were able to implement support for CityJSON with ease, we are certain that it will be easy for other developers to do the same.

Our JSON-based encoding allows practitioners to continue use the CityGML data model, as it is simply an extra encoding; the features not implemented, which are a few, are in our opinion rarely used, are meant to keep the encoding simple, and are well-documented on the website.
For exchanging datasets, but also for creating and editing them, we believe CityJSON offers a more flexible encoding, and the fact that files are more compact (~6$\times$ in practice) is beneficial, especially in a web context.
Since there is open-source software to convert---without loss of information---between the JSON and the GML encodings, one can decide to perform some tasks with CityJSON and some others with the GML encoding.

We believe CityJSON will be useful to the whole community because it will foster the development of (open-source) tools from small programmers and researchers, and it will make it easier for practitioners to exchange and process their datasets.
The development of the CityJSON specifications (and its accompanying software) is open on GitHub, and everyone is welcome to contribute.

%

As future work, we plan to implement a tiling scheme to subdivide large files into different parts, using for instance a quadtree.
We also plan to offer a binary encoding, using for instance Binary JSON (BSON)\footnote{\url{http://bsonspec.org/}}; this would allow us to compress even more the files.
Finally, when the new specifications for CityGML v3 will be released by the OGC, we will study them and modify the CityJSON specifications accordingly, as long as they do not clash with the principles of simplicity and usability that CityJSON is based upon.
We plan on continuing to develop CityJSON to make it as usable as possible in practice, and we invite others to join us and propose new features to add (or to delete, for the sake of simplicity!).

\bibliographystyle{spbasic} 
\bibliography{ref}

\end{document}